# Development and characterization of *Brassica juncea – fruticulosa* introgression lines exhibiting resistance to mustard aphid (*Lipaphis erysimi* Kalt)

Chhaya Atri[†], Bharti Kumar[†], Hitesh Kumar, Sarwan Kumar, Sanjula Sharma and Surinder S Banga[*]

## Abstract

**Background:** Mustard aphid is a major pest of *Brassica* oilseeds. No source for aphid resistance is presently available in *Brassica juncea* . A wild crucifer, *Brassica fruticulosa* is known to be resistant to mustard aphid. An artificially synthesized amphiploid, AD-4 (*B. fruticulosa* × *B. rapa* var. brown *sarson*) was developed for use as a bridge species to transfer *fruticulosa* resistance to *B. juncea*. Using the selfed backcross we could select a large number of lines with resistance to mustard aphid. This paper reports cytogenetic stability of introgression lines, molecular evidence for alien introgression and their reaction to mustard aphid infestation.

**Results:** Majority of introgression lines had expected euploid chromosome number(2n= 36), showed normal meiosis and high pollen grain fertility. Well-distributed and transferable simple-sequence repeats (SSR) markers for all the 18 *B. juncea* chromosomes helped to characterize introgression events. Average proportions of recipient and donor genome in the substitution lines were 49.72 and 35.06%, respectively. Minimum alien parent genome presence (27.29%) was observed in the introgression line, Ad3K-280 . Introgressed genotypes also varied for their resistance responses to mustard aphid infestations under artificial release conditions for two continuous seasons. Some of the test genotypes showed consistent resistant reaction.

**Conclusions:** *B.juncea-fruticulosa* introgression set may prove to be a very powerful breeding tool for aphid resistance related QTL/gene discovery and fine mapping of the desired genes/QTLs to facilitate marker assisted transfer of identified gene(s) for mustard aphid resistance in the background of commercial mustard genotypes.

**Keywords:** Insect resistance, Indian mustard, *Lipaphis erysimi*, Wild crucifer, Wide hybridization

## Background

Indian mustard (*Brassica juncea* L.Czern & Coss.) is an important winter oilseeds crop of India. Mustard aphid, *Lipaphis erysimi* (Kalt.) is one of the most damaging of biotic stresses that confront this crop. It is highly host specific, feeding exclusively on *Brassica* phloem sap. Retarded growth, poor seed formation and low oil content are the prominent manifestations of parasitic feeding and consequent source restrictions in brassica oilseeds [1-3]. During outbreak years; the mustard aphids can cause up to 70% productivity losses. Parthenogenesis and fast growth results in nymphs attaining reproductive age in less than 10 days. Such an enormous propagation rate gets manifested in abnormally high aphid population under favorable conditions. As crucifer specialists, aphids have developed mechanisms to withstand or even alter plant defensive chemicals that normally act as feeding deterrents for generalist herbivores [4]. Present methods of aphid control in mustard are primarily based on synthetic chemical insecticides. These chemicals, besides aggravating environmental pollution, can also be toxic to friendly insects. A resistant cultivar is always a more sustainable and environment-friendly option for managing insect-pests. The development of an insect-resistant cultivar requires a heritable and transferable resistance [5]. Three mechanisms of host plant resistance to insects have been recognized [6]. These are: antibiosis, antixenosis, and tolerance. Antibiosis manifests in the form of toxic chemicals or secondary metabolites that impacts insect biology. Non

* Correspondence: nppbg@pau.edu
[†]Equal contributors
Department of Plant Breeding and Genetics, Punjab Agricultural University, Ludhiana 141 004, India

preference for host plant defines antixenosis resistance. Tolerance is the ability to withstand a given population pressure of insect without significant biological loss. Level of tolerance tends to vary with environment and physiological status of host plant. Glucosinolate-myrosinase system, plant volatiles, phytoalexins, phytoanticipins, sulphur, lectins and numerous other classes of secondary metabolites such as hydroxamic acids, alkaloids, glucosinolates, terpenes and C-6 aldehydes have been reported for their insecticidal activities in various studies [7]. Extensive genetic variability for many defensive traits is known to exist in wild and weedy crucifers [8]. However, several attempts to relate such a variation to heritable resistance in primary gene pool of crop *Brassica* species have not met with any success [9]. This led to many attempts at developing *Brassica* transgenic carrying genes for lectin production that offered higher levels of resistance against *L. erysimi* [10]. Field testing of such a transgenics is still awaited. Other molecular and genomic approaches such as transcript profiling, mutational analysis, over expression, and gene silencing are also now being considered to develop host plant resistance to aphids [11]. Once considered to be the only way to feed the burgeoning world population, unrelenting opposition to genetically modified organisms (GMOs) continue to hamper their greater acceptability in large areas of the globe. An alternative approach to transgenic technology is the exploitation of beneficial genes from wild relatives of crop plants using conventional or genomics aided breeding methods. This is reflected in renewed interest in moving back to wilds to hunt genes of interest equipped with present knowledge of genome and tools of biotechnology that have erased the sexual boundaries for gene transfer.

*B. fruticulosa*, a wild relative of crop Brassicas, has been reported to possess resistance against cabbage aphid, *Brevicoryne brassicae* [12-14] and a higher concentration of lectins was suggested to be the underlying mechanism of resistance in this species [15]. Past investigations of our group allowed the development of an inter specific hybrid between an aphid resistant accession of *B. fruticulosa* and *B. rapa* [16]. Amphiploidy and crossing with *B. juncea* followed by repeated cycles of selfing helped in the synthesis of advanced inter specific derivatives. Laboratory and field evaluation of a large number of lines carrying introgression from *B. fruticulosa* allowed us to demonstrate a significant level of field resistance to aphid infestation [17]. A complete set of introgression lines was developed using single pod descent method following first cycle of backcrossing. This breeding scheme was initiated with the purpose of achieving optimal coverage of *B. fruticulosa* genome in the background of *B. juncea*. Present investigations were undertaken to analyze this set of *B. juncea* introgression lines for cytogenetic stability, aphid resistance and to establish the extent of alien introgression in selected introgression lines using genome specific molecular markers.

## Methods

### Plant materials and population development

Plant materials comprised a set of introgression lines that were developed from the selfing BC1 generation of the cross, *B. juncea*/(*B.fruticulosa*/*B.rapa*). The single pod descent method was followed for their development. The introgression set at the time of evaluation was in BC1S4 and BC1S5 generation(s).

### Field assessment of introgression set for aphid resistance

A set of 533 introgression lines was screened under field conditions during 2009–10, along with the *B. rapa*, *B. juncea*, *B. napus* and *B. fruticulosa* and AD-4 (*B. fruticulosa* × *B.rapa*) amphiploid. The experiment was repeated during 2010–2011 with 221 lines, selected based on the data generated during first year. To supplement aphid population development under natural conditions, aphids were released artificially @ 20 aphids/plant. Five random plants were observed per replication for aphid injury symptoms on a 0–5 scales at flowering and pod formation following the procedure of Bakhetia and Sandhu [18]. (0–5) were assigned to each plant observed. Aphid infestation index (AII) was worked out as per Bakhetia and Bindra [19] to adjudge performance of each test entry. Higher the AII, lower was the level of resistance.

### Genotyping

DNA was isolated using a standard procedure of Doyle and Doyle [20]. Three to four young leaves from four to five week-old plants were collected in a vial and kept in ice until transfer to the laboratory. Molecular studies were carried out using a set of 74 A- and B- genome chromosome specific and transferable SSR primers to document alien introgressions in the B. juncea introgression lines. An automated high-throughput electrophoresis system (CaliperLab Chip GX version 3.0.618.0) was used to separate the PCR product. The data were scored as present “1” or absent “0” for a band at a specific position in a gel with reference to a base-pair ladder. All detectable bands were scored. The length (base pair number) of each fragment in the amplified product was determined concerning the marker ladder.

### Statistical analyses

PAleontological Statistics (PAST) software Version 2.11 [21] was used to construct a principal coordinate analysis for 45 introgression lines as well as for their parents. The estimation of segments from the donor wild parent in the introgression line was carried out with the help of the CSSL Finder v. 0.84 computer program [22]. Means were separated using the least significant difference (LSD) test at probability of P=0.05 with the statistical software OPSTAT.

## Results

### Phenotyping of introgression set for aphid resistance

Five hundred and thirty three introgression lines were tested for their reaction to *L. erysimi* infestation along with the *B. rapa*, *B. juncea*, *B. napus*, *B.fruticulosa* and *B. fruticulosa* × *B. rapa* amphiploid under artificial release conditions during 2009–2010. For the 2010–11 season, a select set of 221 lines was screened for three aphid-related parameters. Data were recorded for aphid population per plant, per cent plant infestation and aphid infestation index. Aphid population/plant ranged from zero to 175 during first year of evaluation. Frequency distribution as depicted in Figure 1 revealed that proportion of plants harboring lower aphid population was higher than the plants with heavy aphid infestation. Trend was almost similar during 2010–2011 and it indicated consistency of the resistance responses over two years on an aphid population basis. The results from the parameter of per cent plants infested were not categorical. During 2009–2010, almost 80% genotypes were infested by the mustard aphids. There seemed to be a mismatch between plant infestation and aphid population. The proportions of plants with heavy infestation were lower during 2010–2011 as compared to the previous year. Frequency distribution data showed two predominant peaks (Figure 1) one for the resistant class another for the susceptible class. The AII ranged from < 1 to 5. Frequency distribution for aphid infestation index was skewed in favour of resistant category (Figure 1). AII for *B. fruticulosa*,amphiploid (*B. fruticulosa* × *B. rapa*) and susceptible check, *B. rapa* cv. BSH

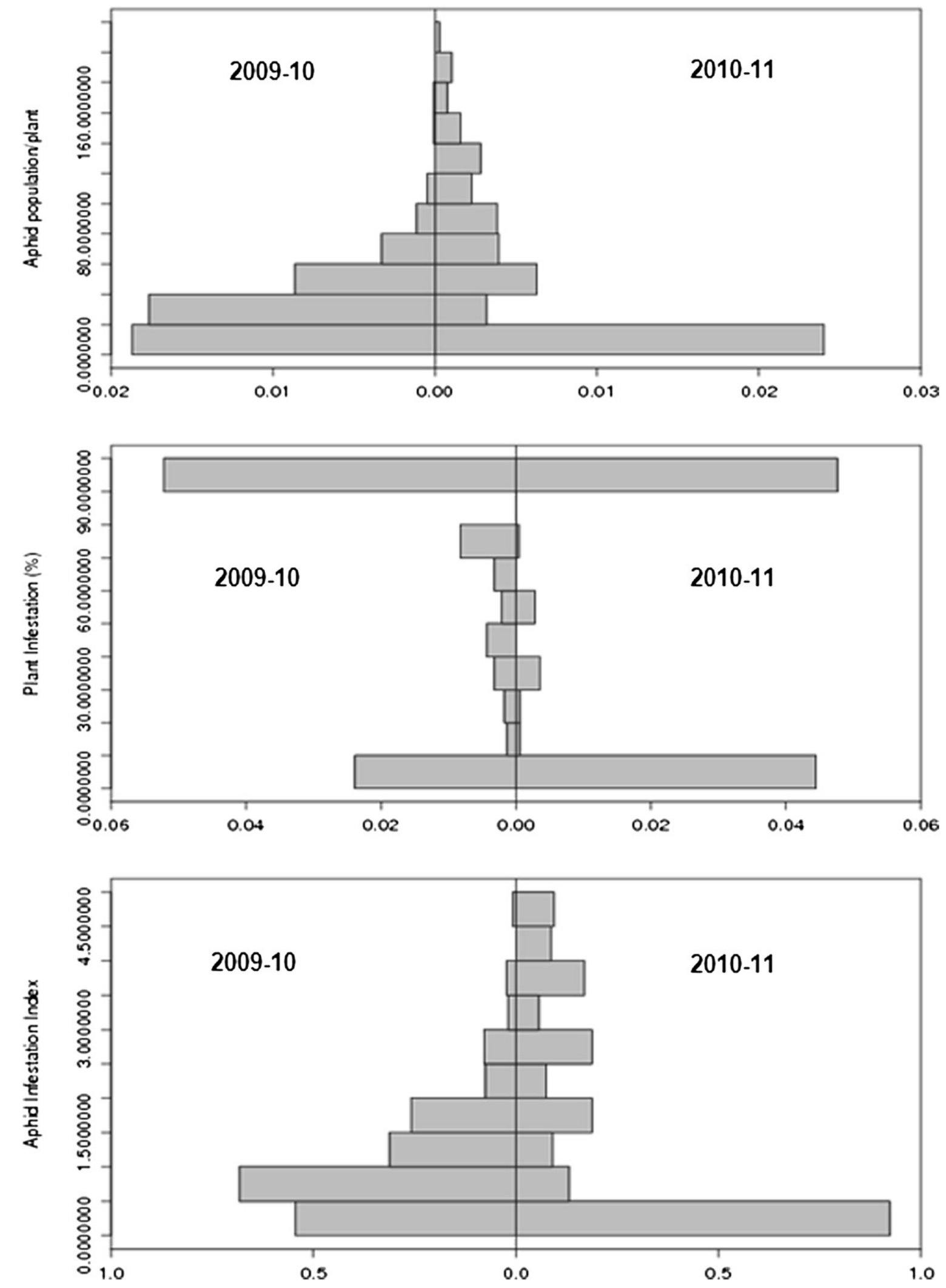

**Figure 1** Frequency histogram for reaction of different *Brassica juncea* introgression lines to mustard aphid infestation.

−1 were 1.0, 1.0, and 2.36 respectively . Overall, the frequency distribution was biased in favour of lower aphid population, low proportion of infested plants in resistant category and consequently, low AII.

Analysis of variance, using year as replication showed significant replication and genotype differences (Table 1). There were a large number of nitrogression lines that showed consistent resistant reaction over two years. The resistant inbred lines harbored an average of 3.0 aphids/plant compared to 96.0 aphids/plant on susceptible lines. Genotypes namely Ad3L-490, Ad3K-284, Ad3L-384, Ad3K-135, Ad3L-391, Ad3L-439, Ad3L-408, Ad3L-491, Ad3L-429 and Ad3L-393 had the lowest score for aphid pop/plant, percent plant infestation and aphid infestation index (Table 2).

### Genomic stability of the introgression lines

Tests for pollen grain viability and meiotic analysis of the introgression lines were conducted to study the genomic stability of the introgression lines in BC1S4 and BC1S5. Pollen grain staining by acetocarmine allowed determination of the pollen viability in a set of 273 introgression lines of the total of available 533 lines during 2009–2010. Pollen grain fertility as inferred from pollen grain stain ability ranged from 0 to 100% over both the crop seasons (Figure 2a). Single generation of selection for fertility during BC1S3 generation in a set of introgression lines helped in achieving significantly enhanced fertility levels in BC1S5 with frequency distribution skewed towards greater pollen grain fertility (Figure 2b). Ad3K-114, Ad3K-379, Ad3K-65, Ad3L-368 and Ad3K-28 had the highest value for pollen grain viability. Meiotic configurations were studied to select introgression lines showing steady reaction to aphid infestation over the years. Most genotypes revealed expected euploid chromosome number (2n=36) with stable 18II during metaphase and 18–18 separation during anaphase (Figure 3a-b).

**Table 1 Analysis of variance for traits related to aphid resistance**

| Source | ANOVA | | | |
|---|---|---|---|---|
| | Aphid Infestation Index | Aphid pop/plant | % plant infestation | Pollen grain Viability (%) |
| **Replications** | 384.067** | 203.091** | 55863.580** | 626.312** |
| **Treatments** | 12.625** | 9.903** | 1713.452** | 108.323** |
| **Error** | 6.852 | 4.319 | 923.983 | 51.494 |
| **C V** | 73.26 | 54.58 | 73.97 | 10.57 |

*Significant at 5% level of significance.
**Significant at 1% level of significance.

### Molecular studies to confirm alien introgression

A set of 45 introgression lines and the donor species, *B. fruticulosa* was used for molecular characterization with known. A and B genome specific SSR primers. The purpose was to confirm *B. fruticulosa* introgression in *B. juncea*. DNA polymorphism generated by 74 transferable SSR primers was utilized for principal coordinate analysis. Principal coordinate analysis (Figure 4) suggested three broad groups, *B. fruticulosa* was the most diverse. In between was a very large group comprising most introgression lines. There was a third loose grouping that comprised six genotypes. Standard *B.juncea* cv. RLC-1 was closest to introgression lines AD3L-368, AD3L-341 and AD4K2-196. Interestingly, these genotype also revealed high pollen grain fertility and normal meiosis as well as the desirable aphid infestation indices 0.0,0.0 and 1.4 respectively. CSSL Finder program selected a subset of 74 SSR markers distributed across eighteen *B. juncea* chromosomes. On this basis, identifying CSSL candidates led to a set of forty five lines (Figure 5). The recipient parent component in the introgression lines varied from 37.32 to 64.21% with the average of 49.72%. The donor parent contribution to the introgression lines ranged from 27.29 to 48.30% (Table 2). Many of the lines revealed heterozygous chromosomal regions. Ad3K-280 had minimum contribution of the alien genome and was resistant to mustard aphid infestation. This line can be a useful source for fine mapping of the gene(s) for aphid resistance.

## Discussion

From point of view of crop evolution, plant resistance to insects has been defined in terms of heritable characteristics that allow plants to withstand insect attack and reproduce. Several factors may constitute resistance. Tolerance and avoidance sometimes are insufficient as their efficacy can be compromised by population load (tolerance) or the multiplicity of crops, allowing aphids may move even to the less preferred host for feeding or oviposition. Antibiosis provides excellent resistance; however, as it exercises maximum pressure on insect, there is always a possibility of evolution of resistant biotypes. Such an eventuality can compromise the stability of resistance. An ideal resistance may result from a combination of all three mechanisms with tolerance showing slightest pressure on the insect to adapt [23]. The results of the feeding preference/choice test had earlier revealed that mustard aphid showed lower preference for *B. fruticulosa* [17]. The antixenosis to feeding in B. fruticulosa is also known for cabbage aphid, *Brevicoryne brassicae* [12,13] and cabbage root fly, *Delia radicum* [24]. High level of resistance in *B. fruticulosa* has been attributed to a combination of antixenosis and antibiosis. Introgressive breeding is a difficult, random and rare process, success of which depends a lot on the existence of homology between donor and recipient genomes. The problem is still complex if one or both species in question are amphiploid, requiring special crossing schemes. Phylogenetic studies in *Brassicaceae* have shown two distinct Brassica lineages namely "Nigra" lineage (B genome) and "Rapa/Oleracea" lineage (A/C genome). *B. fruticulosa* belongs to "Nigra" lineage and was expected to be genetically closer to B genome of *B. juncea* (AABB), the recipient species in the present context. To maximize chances of homoeologous pairing between B genome of *B. juncea* (AABB) and F genome of *B. fruticulosa* (FF), an interspecific amhiploid (AAFF) was first developed following chromosome doubling in the F1 hybrid (AF) between *B. rapa* and *B. fruticulosa*. The resultant amphiploid was then hybridized as a bridging species with *B. juncea*. As visualized, the disomic dosage for A genome in the hybrid (AABF) provided relative genomic stability leading to partial male as well as female fertility in the hybrid; which allowed further backcrossing with the recipient *B. juncea*. That homoeologous pairing and genetic exchanges occurred between recipient, and the donor genome was first evident from excellent variation for aphid resistance in the introgression lines produced. This also emphasized the heritable nature of *B. fruticulosa* resistance.

**Table 2 Phenotyping and genotyping data for a set of selected introgression lines of *Brassica juncea***

| S.No | Introgression Lines | Pollen viability (%) | Aphid pop/ Plant | Plant Infestation (%) | AII | Resistance reaction | % recipient mk | % donor mk | % htz mk |
|---|---|---|---|---|---|---|---|---|---|
| 1 | Ad3K-114 | 98 | 0 | 0 | 0 | R | 43.23 | 48.30 | 5.19 |
| 2 | Ad3K-286 | 90 | 24 | 40 | 0.8 | R | 44.84 | 35.43 | 18.27 |
| 3 | Ad3K-284 | 95 | 0 | 0 | 0 | R | 38.86 | 45.50 | 9.93 |
| 4 | Ad3K-135 | 90 | 0 | 0 | 0 | R | 37.32 | 47.91 | 6.60 |
| 5 | Ad3K-288 | 90 | 80 | 100 | 3 | R | 50.97 | 29.66 | 13.67 |
| 6 | Ad3K-316 | 95 | 65 | 100 | 2.4 | MR | 49.32 | 39.60 | 6.36 |
| 7 | Ad3K-379 | 100 | 0 | 0 | 0 | R | 52.77 | 35.75 | 9.96 |
| 8 | Ad3K-458 | 90 | 24 | 100 | 1 | R | 50.23 | 32.12 | 11.86 |
| 9 | Ad3K-165 | 95 | 0 | 0 | 0 | R | 56.58 | 29.62 | 5.12 |
| 10 | Ad3K-280 | 87 | 0 | 0 | 0 | R | 54.80 | 27.29 | 13.52 |
| 11 | Ad4K2-64 | 89 | 84 | 100 | 3 | S | 48.64 | 34.50 | 10.72 |
| 12 | Ad4K2-77 | 90 | 40 | 100 | 1.4 | R | 45.29 | 34.86 | 10.16 |
| 13 | Ad3K-65 | 98 | 0 | 0 | 0 | R | 52.11 | 34.10 | 9.40 |
| 14 | Ad4K2-185 | 90 | 6 | 40 | 0.4 | R | 64.21 | 30.13 | 2.48 |
| 15 | Ad4K2-196 | 90 | 42 | 100 | 1.4 | R | 50.89 | 36.63 | 5.72 |
| 16 | Ad4K2-63 | 90 | 84 | 100 | 2.6 | MR | 56.81 | 32.27 | 4.52 |
| 17 | Ad4K2-101 | 95 | 146 | 100 | 4.2 | S | 48.34 | 41.23 | 2.64 |
| 18 | Ad3L-421 | 85 | 0 | 0 | 0 | R | 47.43 | 32.62 | 11.27 |
| 19 | Ad3L-408 | 85 | 0 | 0 | 0 | R | 54.75 | 32.67 | 4.67 |
| 20 | $Ad_3L$-341 | 90 | 0 | 0 | 0 | R | 46.34 | 41.00 | 2.15 |
| 21 | Ad3L-354 | 80 | 0 | 0 | 0 | R | 55.63 | 29.03 | 8.46 |
| 22 | Ad3L-368 | 98 | 0 | 0 | 0 | R | 49.16 | 29.60 | 3.67 |
| 23 | Ad3L-373 | 80 | 60 | 100 | 2 | MR | 44.53 | 36.21 | 8.16 |
| 24 | Ad3L-384 | 95 | 0 | 0 | 0 | R | 49.18 | 31.13 | 8.31 |
| 25 | Ad3L-391 | 89 | 0 | 0 | 0 | R | 50.19 | 28.85 | 16.77 |
| 26 | Ad3L-393 | 95 | 0 | 0 | 0 | R | 51.00 | 35.98 | 8.88 |
| 27 | Ad3L-394 | 82 | 0 | 0 | 0 | R | 54.32 | 32.02 | 6.50 |
| 28 | Ad3L-429 | 95 | 0 | 0 | 0 | R | 58.50 | 28.82 | 5.64 |
| 29 | Ad3L-439 | 80 | 0 | 0 | 0 | R | 53.76 | 31.84 | 7.10 |
| 30 | Ad3L-440 | 95 | 0 | 0 | 0 | R | 50.12 | 38.41 | 8.17 |
| 31 | Ad3L-449 | 78 | 0 | 0 | 0 | R | 57.73 | 32.69 | 6.60 |
| 32 | Ad3L-467 | 85 | 0 | 0 | 0 | R | 42.66 | 40.99 | 10.31 |
| 33 | Ad3L-478 | 90 | 36.6 | 100 | 1.3 | R | 58.26 | 31.27 | 4.25 |
| 34 | Ad3L-491 | 80 | 0 | 0 | 0 | R | 48.11 | 39.20 | 10.89 |
| 35 | Ad3L-490 | 85 | 0 | 0 | 0 | R | 44.30 | 41.36 | 8.55 |
| 36 | Ad3L-497 | 85 | 0 | 0 | 0 | R | 51.73 | 29.61 | 12.16 |
| 37 | Ad3L-505 | 95 | 0 | 0 | 0 | R | 43.80 | 35.27 | 11.95 |
| 38 | Ad3L-529 | 80 | 0 | 0 | 0 | R | 47.04 | 34.37 | 11.92 |
| 39 | Ad3L-520 | 88 | 0 | 0 | 0 | R | 45.19 | 35.25 | 10.96 |
| 40 | Ad3L-524 | 90 | 0 | 0 | 0 | R | 50.19 | 33.59 | 9.57 |
| 41 | Ad3L-539 | 85 | 0 | 0 | 0 | R | 47.40 | 43.06 | 2.61 |
| 42 | $Ad4K_1$-6 | 80 | 74 | 100 | 2.4 | MR | 42.19 | 43.84 | 5.65 |

**Table 2 Phenotyping and genotyping data for a set of selected introgression lines of *Brassica juncea*** *(Continued)*

| | | | | | | | | | |
|---|---|---|---|---|---|---|---|---|---|
| 43 | Ad3K-28 | 98 | 0 | 0 | 0 | R | 50.26 | 29.06 | 10.39 |
| 44 | Ad4K2-65 | 90 | 118 | 100 | 3.4 | S | 42.98 | 31.12 | 7.18 |
| 45 | Ad4K2-79 | 85 | 96 | 100 | 3.6 | S | 55.25 | 33.72 | 4.69 |
| 46 | *B. Fruticulosa* | 0 | 0 | 0 | 0 | R | | | |
| | | | | | | **Min** | **37.32** | **27.29** | **2.15** |
| | | | | | | **Max** | **64.21** | **48.30** | **18.27** |
| | | | | | | **Average** | **49.72** | **35.06** | **8.30** |

A total of 533 introgression lines was initially screened for aphid population per plant, per cent plant infestation and aphid infestation index (AII). Varied level of aphid resistance was recorded based on all the three criteria. Frequency distribution showed a greater proportion of plants with a resistant grade. During 2010–11, a select set of 221 lines was screened. Trend was almost similar to that observed for aphid population/plant and aphid infestation index. The frequency distribution was skewed in favor of low proportion of infested plants in resistant category. Response to selection was indicated. There was a reduction of within progeny variation in BC1S5 as compared to BC1S4 and year wise variations in environmental conditions may also be the factors impacting aphid incidence. These emphasize the existence of real and heritable resistance to mustard aphid in the wide cross progenies evaluated. Pink et al. [25] have earlier reported fixation of high levels of resistance to *B. brassicae* in true breeding lines of *B. fruticulosa*. Genotypes namely Ad3L-490, Ad3K-284, Ad3L-384, Ad3K-135,

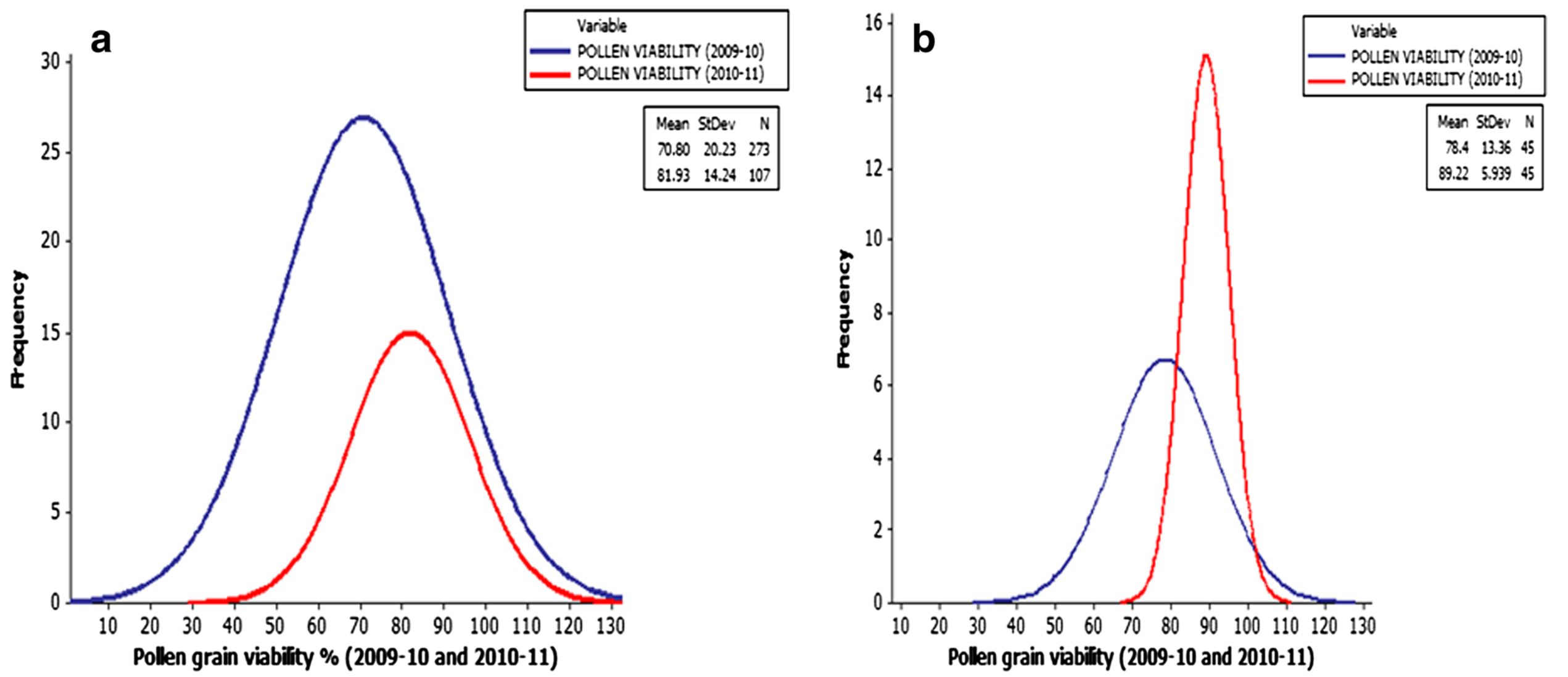

**Figure 2 Frequency distribution for pollen grain fertility. a**) complete introgression lines, **b**) selected sample of introgression lines.

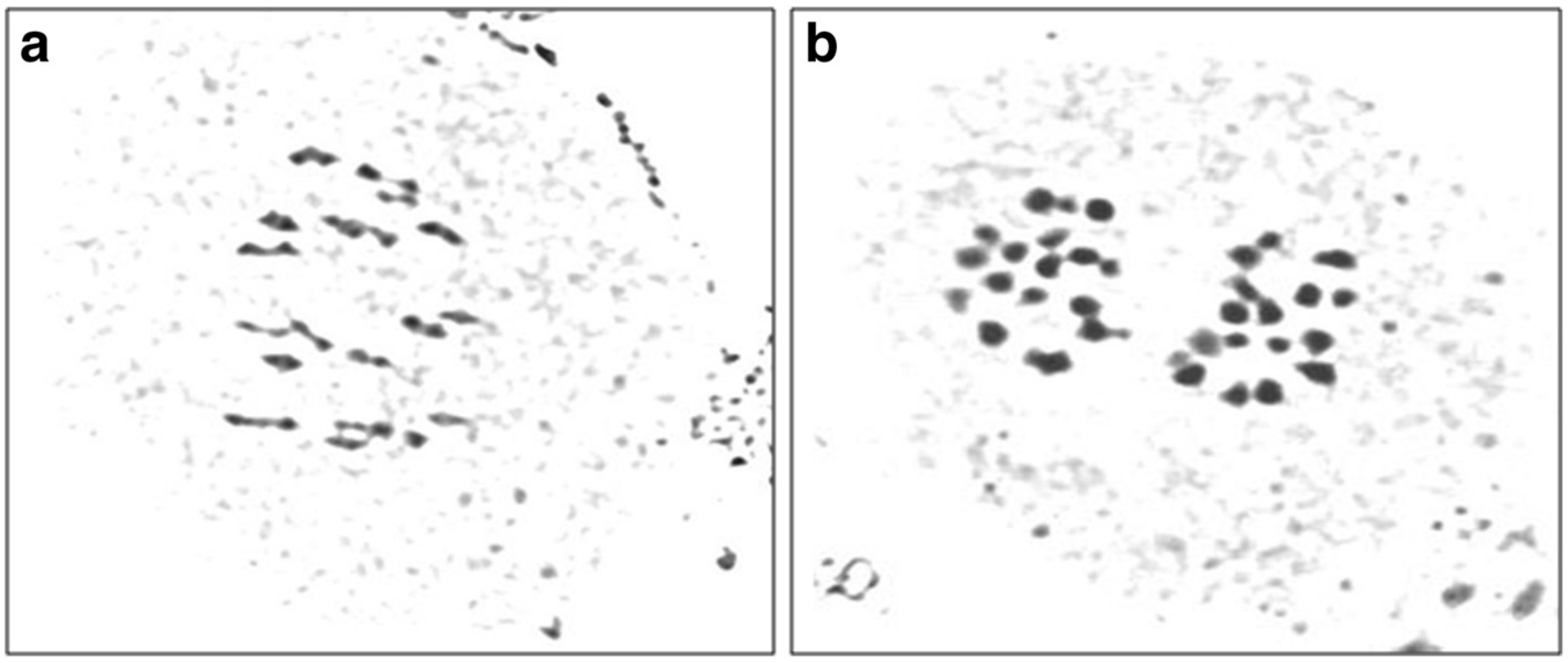

**Figure 3 Meiotic configuration in a stable introgression line of *Brassica juncea*; a**) 18II at metaphase 1, **b**) 18-18 separation at anaphase 1.

Ad3L-391, Ad3L-439, Ad3L-408 Ad3L-491, Ad3L-429 and Ad3L-393 had the lowest score for aphid pop/plant, per cent plant infestation and aphid infestation index. Genomic stability of the introgression lines was apparent from high pollen grain fertility and normal meiosis. Few fully male sterile plants were also identified; they seem likely to have been arisen from interaction of *B. juncea* genome and *B. fruticulosa* cytoplasm. Such manifestation of nucleo-cytoplasmic interactions has been reviewed adequately [26]. There was a conscious attempt to facilitate introgression as well as to improve stability of introgressions and achieve normal euploid chromosome number for *B. juncea*. Partial fertility can result from meiotic irregularities expected in the interspecific hybridization. Molecular characterization of the introgression lines showed excellent alien genome coverage. A resistant genotype, Ad3K-280 had least contribution from donor alien species. This genotype may be especially useful for generating new mapping populations for fine mapping the gene(s) for resistance. A significant proportion of heterozygous loci were also detected. This can result from slower approach to homozygosity for the chromosomes carrying alien introgression or due to chance out crossing. Few more generations of selfing or additional backcrossing with marker-assisted monitoring is likely to help in further reducing the contribution of alien genome to bare minimum required.

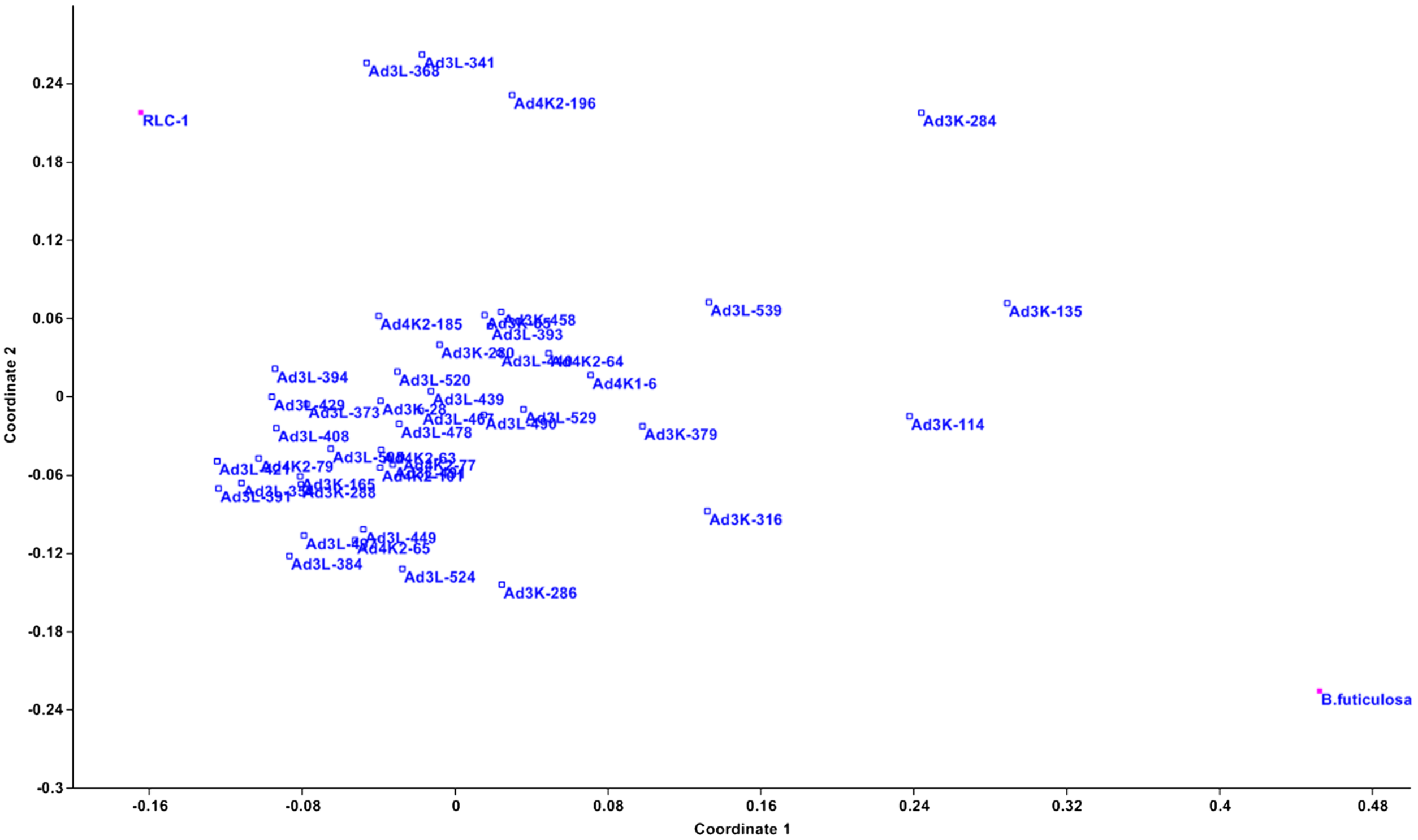

**Figure 4 Principal coordinate analysis showing the genetic diversity among different introgression lines used in the studies.**

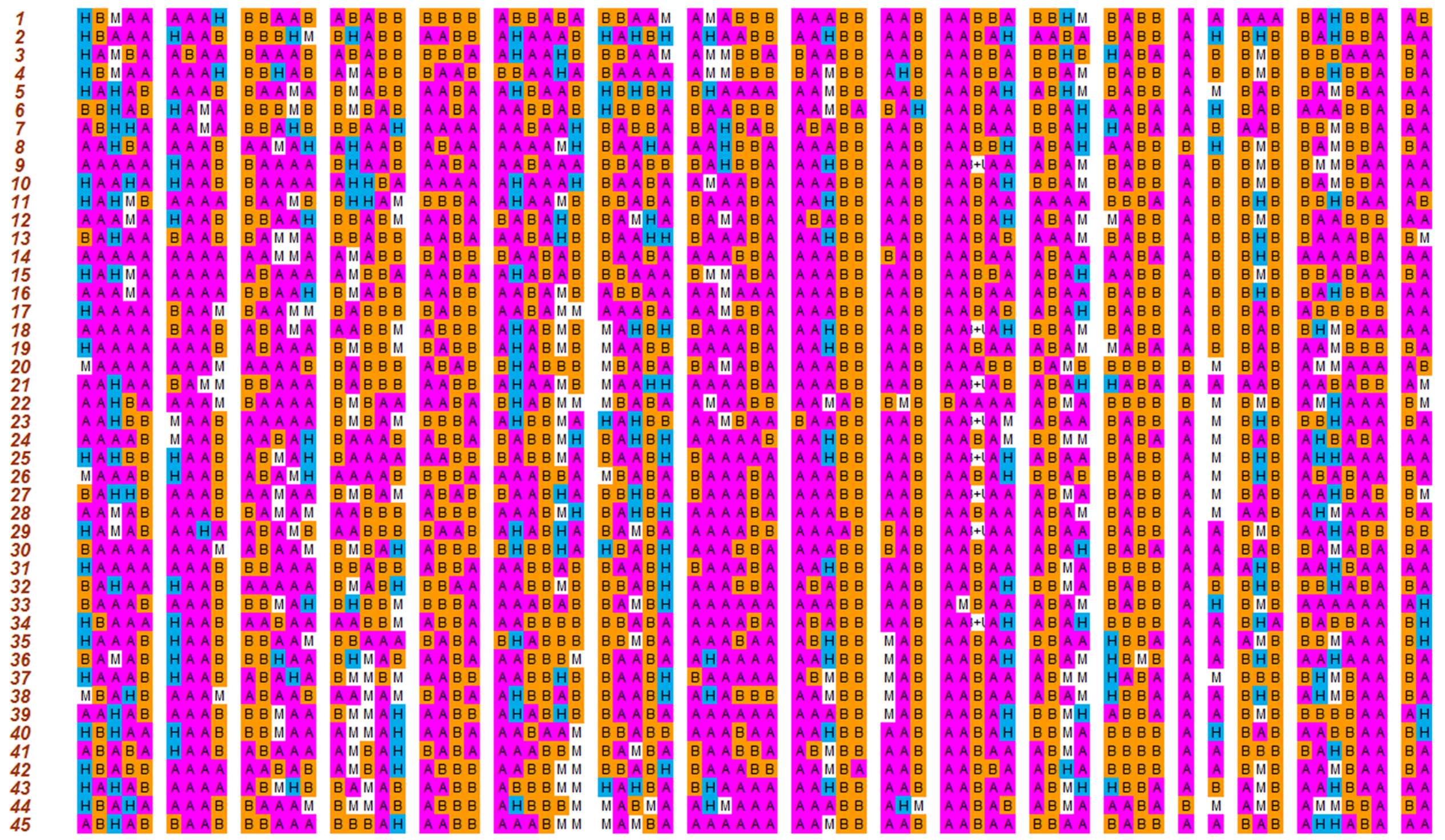

**Figure 5 Graphical genotypes of the selected $BC_1S_4$ lines.** Each row represents a candidate line and each column a linkage Group. H indicates the heterozygous (wild/cultivated) segments, **A** and **B** denotes homozygous regions for cultivated and wild alleles respectively. M indicates missing data.

## Conclusions

Screening of the introgression lines of *B. juncea* with genetic information from *B. fruticulosa* helped in identification of genotypes possessing significantly higher level of resistance to mustard aphid. Genomic stability of introgressions was also established at the population level.*B.juncea-fruticulosa* introgression set may prove to be a very powerful breeding tool for aphid resistance related QTL/gene discovery and fine mapping of the desired genes/QTLs to facilitate marker assisted transfer of identified gene(s) for mustard aphid resistance in the background of commercial mustard genotypes.

### Competing interests

The authors declare that they have no competing interests.

### Authors' contributions

SSB developed the concept, designed experiments, and produced introgression lines. CA associated with the development of introgression lines and conducting molecular investigations; BK assisted in field, cytological and molecular investigations, SK collected data regarding aphid infestation. HK and SS helped in molecular studies CA, SK, BK, HK compiled and analysed the data. SSB interpreted results and wrote the paper. All authors read and approved the final manuscript.

### Acknowledgements

The studies were funded by the ICAR National Professor Project "Broadening the genetic base of Indian mustard (*Brassica juncea*) through alien introgressions and germplasm enhancement" awarded to S. S. Banga.

Received: 15 August 2012 Accepted: 22 November 2012
Published: 27 November 2012